\documentclass[prd,aps,secnumarabic,superscriptaddress,twocolumn]{revtex4}
\usepackage{graphicx}
\usepackage{bm}
\usepackage{amsmath}
\usepackage{amsfonts}
\usepackage{amssymb}
\def\Dot{\!\cdot\!}

\def\ep{\varepsilon}
\def\sl#1{#1\hspace{-7pt}/}
\def\tr{\mathrm{Tr}}
\def\be{\begin{equation}}
\def\ee{\end{equation}}
\def\bea{\begin{eqnarray}}
\def\eea{\end{eqnarray}}
\def\bc{\begin{displaymath}}
\def\ec{\end{displaymath}}
\def\Li2{\mathrm{Li_2}}
\begin{document}
\title{Noncommutative QCD corrections to the gluonic decays of heavy quarkonia }
\author{Alberto Devoto}
\email{Alberto.Devoto@ca.infn.it}
\affiliation{Dipartmento di Fisica, Universit\`a di Cagliari and INFN, Sezione di Cagliari, Cagliari, Italy} 
\author{Stefano Di Chiara}
\email{Stefano.DiChiara@ca.infn.it}
\affiliation{Dipartmento di Fisica, Universit\`a di Cagliari and INFN, Sezione di Cagliari, Cagliari, Italy}
\author{Wayne W. Repko}
\email{repko@pa.msu.edu} 
\affiliation{Department of Physics and Astronomy, Michigan State University, East Lansing, MI 48824}

\date{\today}

\begin{abstract}
We compute the Noncommutative QCD (NCQCD) contributions to the three gluon decay modes of heavy quarkonia. For triplet quarkonia (ortho-quarkonia), the NCQCD correction to the QCD  three gluon decay mode, like the standard model contribution, is infrared finite. In the case of singlet quarkonia (para-quarkonia), whose QCD three gluon decay mode has infrared singularities which are removed using one-loop corrections to the two gluon mode, we find that NCQCD contribution is also infrared finite. The calculations are performed in the weak binding limit and do not require the introduction of additional effective couplings.
\end{abstract}
\maketitle

\section{Introduction}

Efforts to explore the physical implications of field theories formulated on noncommutative spaces have increased recently \cite{apps,hkm} due to developments in string theories \cite{cds,dh,sw}, which suggest that noncommutative field theories are well defined quantum theories \cite{dh}. In noncommutative geometry, the coordinates $x^{\mu}$ obey the commutation relations
\begin{equation}
\left[x^{\mu},x^{\nu}\right]= i\theta^{\mu\nu}\,,
\end{equation}
where $\theta^{\mu\nu}=-\theta^{\nu\mu}$. A noncommutative version of an ordinary field theory can be obtained by replacing all ordinary products with Moyal $\star$ products defined by
\begin{equation}
(f\star g)(x)=\left.\exp\left(\textstyle\frac{1}{2}\theta^{\mu\nu}\partial_{x^{\mu}}
\partial_{y^{\nu}}\right)f(x)g(y)\right|_{x=y}\,.
\end{equation}
Here, we use the generalization of the QCD Lagrangian
\begin{equation}\label{lag}
\mathcal{L}=\bar{\psi}\star\sl{D}\,\psi - m\bar{\psi}\star\psi - \frac{1}{2g^2}\tr\left(F_{\mu\nu}\star F^{\mu\nu}\right)\,,
\end{equation}
where
\begin{eqnarray}
D_{\mu}\psi & = & \partial_{\mu}\psi - iA_{\mu}\star\psi\,, \\
F_{\mu\nu}& = & \partial_{\mu}A_{\nu}-\partial_{\nu}A_{\mu}-i\left[A_{\mu},A_{\nu}\right]_{\star}\,,
\end{eqnarray}
and
\begin{equation}
\left[A_{\mu},A_{\nu}\right]_{\star}=A_{\mu}\star A_{\nu}-A_{\nu}\star A_{\mu}\,.
\end{equation}
When supplemented with a gauge fixing term, including a ghost contribution, the Lagrangian Eq.\,(\ref{lag}) can be used to obtain a set of vertices and Feynman rules for perturbative calculations \cite{aa,bs}. 

\section{NCQCD Corrections to the Lifetimes of heavy Ortho and Para Quarkonia }

In standard model QCD, the hadronic contributions to the widths of ground state quarkonia are attributed to gluonic decays. The pseudoscalar states, para-quarkonia, decay predominantly into two gluons while the vector states, ortho-quarkonia, being color singlet spin one states, must decay into three gluons. Unlike para-positronium, which cannot decay into three photons in ordinary QED due to charge conjugation symmetry, para-quarkonium can decay into three gluons. However, the three gluon mode is infrared singular and one-loop corrections to the two gluon mode must be included to obtain a finite contribution to the hadronic width \cite{etac}. 

By assuming that quarkonia are weakly bound it is possible to calculate the NCQCD correction to the three gluon lifetimes by computing the annihilation amplitudes for a noninteracting quark and antiquark at rest and supplying a factor of the square of the bound state wave function at the origin, $|\psi(0)|^2$, to account for the leading binding effect. There is no need to devise an effective interaction as in the case of the pion decay into three photons \cite{gl,CDR}. The NCQCD amplitudes contributing to the three gluon corrections were calculated using the Feynman rules of Ref.\,\cite{aa,bs}. These rules contain contributions involving $\theta_{\mu\nu}$ of the form $k_1^{\mu}\theta_{\mu\nu}k_2^{\nu}$, where $k_1$ and $k_2$ are the momenta of two gluons. To ensure that the unitarity conditions $\theta_{\mu\nu}\theta^{\mu\nu}>0$ and $\ep_{\mu\nu\lambda\rho}\theta^{\mu\nu}\theta^{\lambda\rho}=0$ are satisfied, we take $\theta_{0k}=-\theta_{k0}=0$ and write $\theta_{ij}$ as
\begin{equation} \label{phi}
\theta_{ij}=\frac{1}{\Lambda^2_{NC}}\varepsilon_{ijk}\theta_k,
\end{equation}
where $\theta_k$ is a unit vector and $\Lambda_{NC}$ is the noncommutativity scale. We then have $k_1^{\mu}\theta_{\mu\nu}k_2^{\nu}=\theta\Dot(\vec{k}_1\times\vec{k}_2)/\Lambda_{NC}^2$.

\begin{figure}[h]
\hfill\includegraphics[height=1.0in]{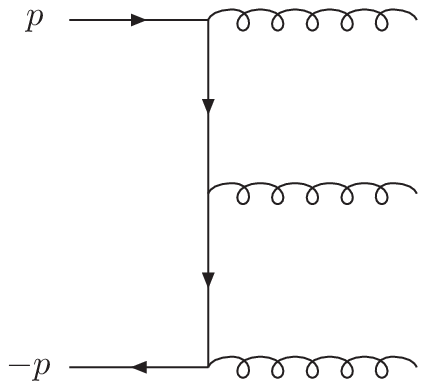}
\hfill\includegraphics[height=1.0in]{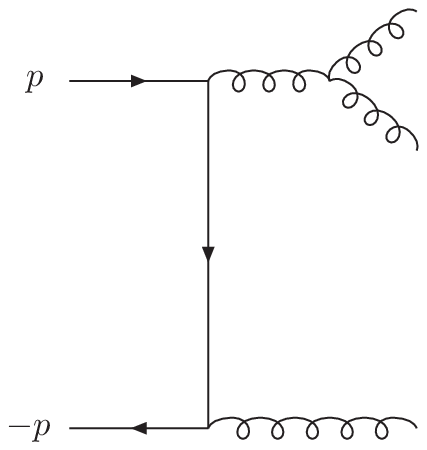}\hspace*{0pt\hfill}
\caption{The left diagram is illustrative of the class of contributions to the three gluon decay of ortho-quarkonium and the right diagram is one of the class of contributions to the para-quarkonium three gluon decay. \label{diags}}
\end{figure}

The amplitudes to be calculated are illustrated in Fig.\,\ref{diags}. Diagrams with the three gluon final state connected to the quark line by a single virtual gluon do not contribute.
We computed the threshold helicity amplitudes with the aid of the symbolic manipulation program FORM \cite{form} and simplified the results using Mathematica \cite{math}. In what follows, we consider quarkonium in the ``non-relativistic'' approximation, taking the quark (and the
anti-quark) 4-momentum $p^{\mu}$ to be $p \equiv (m,0,0,0)$. The gluon 4-momenta are labelled $k_i$, with $i=1,2,3$.

All the squared amplitudes can be expressed as functions of
the parameters $s,t,u$ given by
\begin{equation}
s \equiv (k_1+k_2)^2, \quad
t \equiv (k_2+k_3)^2, \quad
u \equiv (k_3+k_1)^2\,
\end{equation}
which satisfy the identity
\begin{equation}
s+t+u = 4m^2.
\end{equation}
In terms of these variables the squared matrix element
summed over the gluon helicities is, for
the triplet state
\begin{eqnarray} \label{oampsq}
|^3{\cal M}|^2 & = &\frac{128}{27} g^6(5+4\sin^2\varphi)(s+t+u)  \\ 
& \times &\frac{s^2 \left( t+u \right)^2 + t^2 \left( u+s \right)^2 
+ u^2 \left( s+t \right)^2}{\left( s+t \right)^2
\left( t+u \right)^2 \left( u+s \right)^2}\,, \nonumber
\end{eqnarray}
and for the singlet state
\begin{eqnarray}
|^1{\cal M}|^2 & = &\frac{64}{9} g^6 (9 - 4 \sin^2\varphi )(st+tu+us)^2 \\
& \times&\frac{s^4+t^4+u^4+ \left( s+t+u \right)^4 }
{stu \left( s+t \right)^2 \left( t+u \right)^2 \left( u+s \right)^2}\,, \nonumber
\end{eqnarray}
where
\begin{equation} 
\varphi = \frac {1}{2} k_1 ^{\mu} \theta _{\mu \nu} k_2 ^{\nu}\,.
\end{equation}

The integration over phase space for either state is straightforward and the expression for the width in terms of the scaled variables $x=s/4m^2,\;y=t/4m^2$ takes the form
\begin{equation} \label{dGdcos}
\frac {d \Gamma_{Q \rightarrow 3g}}{d \cos \delta}
=\frac {|\psi(0)|^2}{192(2\pi)^3} \int\limits_0^1 dx \,
\int\limits_0^{1-x} dy \, |{\cal M}_{q \bar{q} \rightarrow 3g}|^2\,.
\end{equation}
The variable $\varphi$ can be expressed in terms of $x$, $y$ and the dimensionless scale parameter $z=m^2/\Lambda_{NC}^2$ as
\begin{equation} \label{phi1}
\varphi=\frac{1}{2}|\vec{k_1}\! \times \! \vec{k_2}| |\vec{\theta}| \cos \delta \,
=\sqrt{xy(1-x-y)} z \cos \delta\,.
\end{equation}
\subsection{Ortho-quarkonium}

The ortho-quarkonium decay width can be separated into two terms, $\Gamma_{oQ \rightarrow 3g}^{QCD}$ and $\Gamma_{oQ \rightarrow 3g}^{NCQCD}$.The first, which is independent of $z$, gives, after completing the phase space integration, the standard QCD result \cite{close}
\begin{equation} \label{ogamqcd}
\Gamma_{oQ \rightarrow 3g}^{QCD} = \frac {40} {81} \alpha_s^3
(\pi^2 - 9) \frac {|\psi(0)|^2} {m^2}.
\end{equation}

\begin{figure}[h]
\includegraphics[height=1.6in]{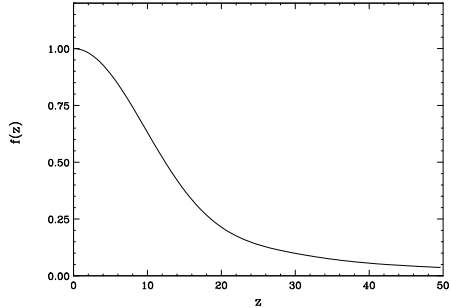}
\caption{The function $f(z)$ is appearing in Eq.\,(\ref{ogamtot}) is plotted.  \label{fortho}}
\end{figure}

Using the symmetry of the integrand with respect to the variables
$s$, $t$ and $u$ (or, equivalently, with respect to the 4-vectors $k_1$, $k_2$
and $k_3$) the ($z$ dependent) NCQCD contribution  can be written
\begin{eqnarray}\label{dgamncqcd}
\frac {d\Gamma_{oQ \rightarrow 3g}^{NCQCD}}{d \cos \delta}
& = &\frac {16}{27} \alpha_s^3 \frac {|\psi(0)|^2}{m^2} \int_0^1 dx \,\int_0^{1-x} dy \\ 
& \times &  
\frac {x^2 \sin^2 \left( \sqrt{xy(1-x-y)} z \cos \delta \right)}
{(x+y)^2 (1-y)^2}\,. \nonumber
\end{eqnarray}
Integration over $d \cos\delta $ ($-1 \le \cos\delta \le 1$) gives
\begin{eqnarray} \label{gamncqcd}
\Gamma_{oQ \rightarrow 3g}^{NCQCD}&=&\frac {16}{27} \alpha_s^3 \frac {|\psi(0)|^2}{m^2}\! \int_0^1\hspace{-8pt} dx \int_0^{1-x}\hspace{-16pt}dy \frac{x^2}{(x+y)^2(1-y^2)}\nonumber\\
&  & \times\left[1 -\frac{\sin\left(2z\sqrt{xy(1-x-y)}\right)}{2z\sqrt{xy(1-x-y)}} \right]. 
\end{eqnarray} 
Due to the presence of the square root in the argument of the sine function in Eq.\,(\ref{gamncqcd}) it is not possible to perform the integration analytically. However, rather than simply keeping the leading term in $z$, we investigated the behavior of the correction to all orders in $z$. The result, when combined with Eq.\,(\ref{ogamqcd}), has the form
\begin{eqnarray} \label{ogamtot}
\Gamma_{oQ \rightarrow 3g}&=&\frac{8}{81}\alpha_s^3\frac {|\psi(0)|^2}{m^2}\left(\mbox{\rule{0pt}{14pt}}5(\pi^2-9)\right. \nonumber \\
&+&\left.\frac{2}{9}(385-39\pi^2)z^2f(z)\right)\,,
\end{eqnarray}
where the behavior of $f(z)$ is illustrated in Fig.\,\ref{fortho}.

\subsection{Para-quarkonium}

The evaluation of the QCD contribution to Eq.\,(\ref{dGdcos}) in the case of para-quarkonium is complicated by the existence of infrared divergences. By requiring the variables $x$ and $y$ to satisfy $x,y\geq\ep$ and using the symmetry of the integrand, the QCD contribution can be written
\begin{eqnarray} \label{pgamir}
\Gamma_{pQ \rightarrow 3g}^{QCD} & = & \frac{4}{3}\alpha_s^3\frac{|\psi(0)|^2}{m^2}\int_{\ep}^{1-2\ep}\hspace{-8pt}dx\int_{\ep}^{1-\ep-x}\hspace{-12pt}dy \\
& \times &\!\frac{(1+3x^4)\left(x^2-(x+y)(1-y)\right)^2}{xy(1-x-y)(1-x)^2(1-y)^2(x+y)^2}\,, \nonumber
\end{eqnarray}
which can be integrated to give
\begin{eqnarray} \label{pgamqcd}
\Gamma_{pQ \rightarrow 3g}^{QCD} & = &\alpha_s^3\frac{|\psi(0)|^2}{3m^2}\left[152-11\pi^2\right. \nonumber \\
& + &\left.4\log(\ep)\left(11+6\log(\ep)\right)\right]\,.
\end{eqnarray}
The infrared behavior of $\Gamma_{pQ \rightarrow 3g}^{QCD}$ exhibited in Eq.\,(\ref{pgamqcd}) must be combined with the one-loop corrections to the two gluon decay to obtain a finite correction to the hadronic width \cite{etac}.

\begin{figure}[h]
\includegraphics[height=1.6in]{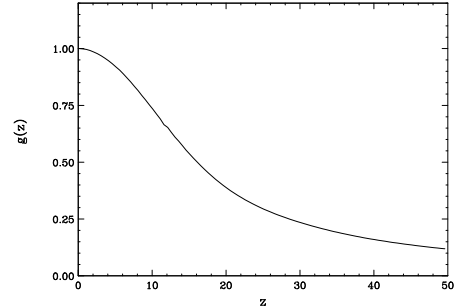}
\caption{The function $g(z)$ is appearing in Eq.\,(\ref{pgamncqcd}) is plotted.  \label{gpara}}
\end{figure}

The evaluation of the non-commutative contribution involves an additional factor of $-4\sin^2\varphi$, which may be handled as in the ortho-quarkonium case to obtain
\begin{eqnarray}\label{ncpgamir}
\Gamma_{pQ \rightarrow 3g}^{NCQCD}& = & -\frac{8}{27}\alpha_s^3\frac{|\psi(0)|^2}{m^2}\int_{\ep}^{1-2\ep}\hspace{-12pt}dx\int_{\ep}^{1-\ep-x}\hspace{-16pt}dy \nonumber\\
& \times &
\frac{(1+3x^4)\left(x^2-(x+y)(1-y)\right)^2}{xy(1-x-y)(1-x)^2(1-y)^2(x+y)^2}\nonumber \\ 
& \times &\left[1 -\frac{\sin\left(2z\sqrt{xy(1-x-y)}\right)}{2z\sqrt{xy(1-x-y)}} \right].
\end{eqnarray}
With the introduction of $\sin^2\varphi$, the integrand of Eq.\,(\ref{ncpgamir}) is no longer singular when $xy(1-x-y)\to 0$ and we may complete the integration with $\ep=0$. The NCQCD contribution then can be written
\begin{equation}\label{pgamncqcd}
\Gamma_{pQ \rightarrow 3g}^{NCQCD}=-\frac{4}{81}\alpha_s^3\frac{|\psi(0)|^2}{m^2}(37\pi^2-\textstyle\frac{5437}{15}) z^2g(z)\,,
\end{equation}
where $g(z)$ is shown in Fig.\,\ref{gpara}.

\section{Discussion and Conclusions}

The inclusion of NCQCD corrections to the three gluon decay widths of ortho- and para-quarkonium does not  change the magnitudes of their hadronic widths substantially for scales $\Lambda_{NC}$ of order $1$ Tev. That said, the NCQCD results are interesting in the sense of what they imply about the consistency of perturbative calculations in these models. 

For ortho-quarkonium, the two gluon decay is forbidden by Yang's theorem. Hence the three gluon decay cannot have any soft gluon singularities because there are no two gluon one-loop corrections available to cancel them. The absence of soft gluon singularities, which is, of course, a feature of the QCD three gluon decay width, persists when the small (positive) NCQCD correction is included. This is guaranteed by the form of the squared amplitude, Eq.\,(\ref{oampsq}).

The situation with para-quarkonium is somewhat more involved because, unlike para-positronium, which can only decay into two photons, both two gluon and three gluon decays are allowed in QCD. In this case, the three gluon decay has infrared singularities, which can be combined with the one-loop corrections to the two gluon decay to obtain a finite contribution to the hadronic width \cite{etac,ratio}. Here, too, the NCQCD correction is infrared finite, but only as a result of a cancellation provided by the NCQCD effective coupling. Interestingly, this correction is negative, and, while small for realistic values of $z=m^2/\Lambda_{NC}^2$, it cannot change the sign of the total width for any value of $z$. 

In summary, we are led to the conclusion that it is not necessary to invoke the smallness of $z=m^2/\Lambda_{NC}^2$ to obtain a sensible NCQCD correction to the hadronic decays of quarkonia. In principle, should we be presented with a fourth generation of very heavy quarks with lifetimes sufficiently long to produce quarkonia, corrections to their decay widths associated with non-commutative geometry could be calculated consistently in the sense that they are finite for any value of $z$.

\begin{acknowledgements}
One of us (WWR) wishes to thank INFN, Sezione di Cagliari and Dipartmento di Fisica, Universit\`a di Cagliari for support. This research was supported in part by the National Science Foundation under Grant PHY-02744789 and by M.I.U.R. (Ministero dell'Istruzione, dell'Universit\`a e della
Ricerca) under Cofinanziamento PRIN 2001.
\end{acknowledgements}


\begin{thebibliography}{99}
\bibitem{ratio} More precisely, one can unambiguously compute the ratio of the QCD radiatively corrected two gluon to two photon widths \cite{etac}.
\end{thebibliography}
\end{document}